\documentclass[12pt]{amsart}
\usepackage[left=15mm, right=15mm, top=15mm, bottom=15mm]{geometry}

\usepackage{float}
\usepackage[font=small,labelsep=period, figurename={Рис.}]{caption}

\usepackage[english]{babel}
\usepackage[T1]{fontenc}
\usepackage[utf8]{inputenc}

\theoremstyle{plain}
    \newtheorem{theorem}{Theorem}[section]

\theoremstyle{definition}

\usepackage[
    draft = false,
    unicode = true,
    colorlinks = true,
    allcolors = blue,
    hyperfootnotes = true,
    citecolor = red
]{hyperref}
\usepackage{amsfonts,amsmath,amssymb,amscd,amsthm,url}
\usepackage[inline]{enumitem}
\usepackage{graphicx,epsf,afterpage, wrapfig}
\usepackage{soul}
\usepackage{multicol}
\usepackage{array}
\usepackage{braket}
\usepackage{epigraph}
\usepackage{rotating, floatflt}
\usepackage{xstring} % Пакет для надежной работы со строками

\usepackage[
backend=biber,
style=numeric-comp,
sorting=none,
giveninits=true
]{biblatex}
\usepackage{csquotes}

\usepackage{xcolor}

\usepackage{ulem}

\usepackage{tikz}
\usetikzlibrary{positioning, calc, intersections, through, arrows, matrix, chains, math}
\usetikzlibrary{arrows.meta}
\usetikzlibrary{shadows}
\usetikzlibrary{graphs}
\usetikzlibrary{graphs.standard}
\usetikzlibrary{patterns}
\usetikzlibrary{decorations.pathreplacing,calligraphy,backgrounds}
\DeclareFieldFormat{usera}{\href{https://arxiv.org/abs/#1}{\texttt{arXiv:#1}}}

\renewbibmacro*{finentry}{\printfield{usera}\newunit\finentry}

\renewbibmacro{in:}{%
  \iffieldundef{usera}
    {\printtext{\bibstring{in}\intitlepunct}}
    {}%
}

\newcommand\norm[1]{\ensuremath{\left\lVert#1\right\rVert}}
\newcommand\abs[1]{\ensuremath{\left\lvert#1\right\rvert}}

\newcommand{\Bin}{\mathrm{Bin}}

\DeclareMathOperator{\Expect}{\mathbb{E}}

\newcommand{\Hbf}{{\ensuremath{\mathbf{H}}}}
\newcommand{\Ibf}{{\ensuremath{\mathbf{I}}}}

\newcommand{\Nbf}{{\ensuremath{\mathbf{N}}}}

\newcommand{\Ubf}{{\ensuremath{\mathbf{U}}}}

\newcommand{\Xbf}{{\ensuremath{\mathbf{X}}}}
\newcommand{\Ybf}{{\ensuremath{\mathbf{Y}}}}

\newcommand{\R}{\ensuremath{\mathbb{R}}}

\renewcommand{\geq}{\geqslant}

\newcounter{mcnt}

\newcounter{wordcnt}

\begin{document}

\title[Quantum Bernstein--Kolmogorov--Arnold approximation]{Quantum Bernstein--Kolmogorov--Arnold approximation}

\author{Sviatoslav V. Dzhenzher}

\begin{abstract}
    This paper addresses the expressivity and representation gap in Quantum Neural Networks by establishing a quantum analogue of the classical Kolmogorov--Arnold representation theorem.
    We propose a constructive approximation framework over fixed input coherent states, where the dependence of the function argument is embedded entirely in coordinate qubits, while the target function dictates a single outer unitary evolution.
    By applying a Sprecher-type representation to the rotation angle of the state, we demonstrate that the multidimensional operator factorises into \(2d+1\) commuting universal inner observables.
    We prove that the resulting measurement probability precisely reproduces the multidimensional Bernstein polynomial of the target function.
    For any continuous multivariate function, the joint quantum state converges uniformly to the target output register, linking classical superposition theory with probabilistic and coherent quantum circuits.
\end{abstract}

\thanks{\hspace{-5mm}
S.\,V. Dzhenzher: sdjenjer@yandex.ru. orcid: 0009-0008-3513-4312
\\
% V.\,Zh. Sakbaev: fumi2003@mail.ru. orcid: 0000-0001-8349-1738
% \\
% V.\,Zh. Sakbaev: Keldysh Institute of Applied Mathematics of Russian Academy of Sciences 125047, Miusskaya pl. 4, Moscow, Russia
% \\
% All authors:
Moscow Institute of Physics and Technology 141701, Institutskii lane 9, Dolgoprudny, Russia}

\maketitle
\thispagestyle{empty}


\vspace{3mm}

\section{Introduction}\label{s:intro}

This work, like many others, originally arose from Hilbert's 13th problem \cite{wiki-Hilb-13-problem, Morris-2021-hilbert13}.
We all know that the roots of a polynomial can be expressed as continuous functions of the coefficients of the polynomial.
Hilbert asked whether any multivariate continuous function can be expressed as the superposition of continuous functions of two variables.
After five decades, Kolmogorov and Arnold showed \cite{Kolmogorov56, Arnold57, Kolmogorov57} that any continuous multivariate function can be expressed as a superposition of univariate functions and addition.
Nowadays, their result, known as the Kolmogorov--Arnold representation theorem (and sometimes referred to as the Kolmogorov Superposition Theorem), can be formulated as follows: \emph{for any integer $d > 1$ and any rationally independent \(\gamma_1,\ldots,\gamma_d\), there exist $2d+1$ ``inner'' continuous functions \(\varphi_0,\ldots,\varphi_{2d}\colon[\,0,1\,]\to[\,0,1\,]\), such that for any ``target'' continuous function \(f\colon[\,0,1\,]^d\to\R\), there exists an ``outer'' uniformly continuous function \(g\colon\R\to\R\), such that for any \(x_1,\ldots,x_d\in[\,0,1\,]\),}
\[
    f(x_1,\ldots,x_d) = \sum_{q=0}^{2d} g\left(\sum_{p=1}^d \gamma_p\varphi_q(x_p)\right).
\]
It took another three decades \cite{HechtNielsen87, Kurkova91, Maiorov-Pinkus} to notice that the Kolmogorov--Arnold representation theorem is relevant in the field of expressivity and Neural Networks, and yet another three decades to use it for the~optimisation of Convolutional Neural Networks \cite{KAN, KAN2, CKAN}.

In the parallel research field of quantum computing, the problem of expressivity is also well known.
For example, in \cite{FLS-prob-quant-Stone-Weier}, the representation of a function is given in terms of the probability of measurement.
The representation is provided in terms of the Stone--Weierstrass approximation, which implicitly gives the Bernstein polynomials.
See, for example, \cite{NacuPeres-fast-bern}, for why the Bernstein polynomial approximations are so effective.
We will presume that our input state is encoded in the coherent states \(\sqrt{1-x_p}\ket{0}+\sqrt{x_p}\ket{1}\), analogously to \cite{DJR-quant-rand, JZS-quant-bern}.
Our approach is close to that of \cite{JZS-quant-bern} in approximating not only the probability but also the input coherent state itself.
As in \cite{Hoch-quant-bern}, we presume that we have access to a sufficiently large number of copies of input coherent states.
So, generally, we try to approximate a coherent state with a unitary evolution of a known input state, which is similar to \cite{Perez-univer-appx}.

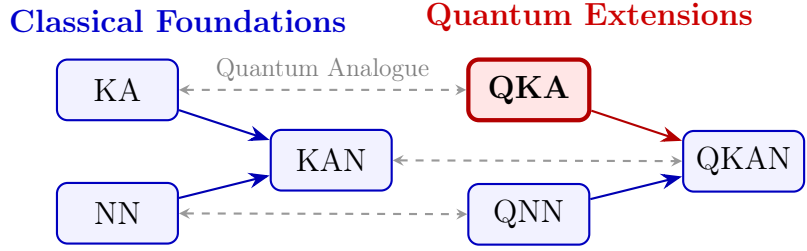
\begin{figure}
    \centering
    \begin{tikzpicture}[
        node distance = 0.8cm and 1.2cm,
        box/.style = {rectangle, draw=blue!75!black, fill=blue!5!white, thick, rounded corners, minimum width=1.6cm, minimum height=0.8cm, align=center},
        gapbox/.style = {rectangle, draw=red!70!black, fill=red!10!white, ultra thick, rounded corners, minimum width=1.6cm, minimum height=0.8cm, align=center},
        groupbox/.style = {draw=gray!30, fill=gray!5, dashed, rounded corners, inner sep=0.3cm}
    ]
        % --- CLASSICAL TREE (LEFT) ---
        \node[box] (ka) {KA};
        \node[box] (nn) [below=of ka] {NN};
        \node[box] (kan) [below right=0.1cm and 1.2cm of ka] {KAN};

        \draw[-{Stealth[scale=1.2]}, thick, blue!70!black] (ka) -- (kan);
        \draw[-{Stealth[scale=1.2]}, thick, blue!70!black] (nn) -- (kan);

        % --- QUANTUM TREE (RIGHT) ---
        \node[gapbox] (qka) [right=3.8cm of ka] {\textbf{QKA}};
        \node[box] (qnn) [below=of qka] {QNN};
        \node[box] (qkan) [below right=0.1cm and 1.2cm of qka] {QKAN};

        \draw[-{Stealth[scale=1.2]}, thick, red!70!black] (qka) -- (qkan);
        \draw[-{Stealth[scale=1.2]}, thick, blue!70!black] (qnn) -- (qkan);
        
        % --- HORIZONTAL CONCEPTUAL CONNECTIONS ---
        \draw[dashed, <->, >=stealth, thick, gray!80] (ka) -- (qka) node[midway, above, scale=0.8, text=gray] {Quantum Analogue};
        \draw[dashed, <->, >=stealth, thick, gray!80] (nn) -- (qnn);
        \draw[dashed, <->, >=stealth, thick, gray!80] (kan) -- (qkan);

        % Labels for Trees
        \node[above=0.2cm of ka, xshift=0.8cm, text=blue!80!black, font=\bfseries] {Classical Foundations};
        \node[above=0.2cm of qka, xshift=0.8cm, text=red!80!black, font=\bfseries] {Quantum Extensions};

    \end{tikzpicture}
    \caption{The knowledge gap in representation theory showing the missing quantum analogues of the Kolmogorov--Arnold representation theorem. Here KA, NN, and KAN mean the Kolmogorov--Arnold representation theorem, Neural Networks, and Kolmogorov--Arnold Networks, respectively; and QKA, QNN, and QKAN mean their quantum analogues}
    \label{fig:qka-gap}
\end{figure}

The main difference between the approaches mentioned above and the approach in this paper is that we try to obtain our unitary evolution in the Kolmogorov--Arnold setting, with fixed ``inner'' operators and an ``outer'' evolution depending on a ``target'' function.
The ideas of applying the Kolmogorov--Arnold representation theorem to the quantum setting and, particularly, to Quantum Neural Networks \cite{QNN-convolut, QNN-deep}, are known as Quantum Kolmogorov--Arnold Networks and are described in many papers \cite{QKAN-china, QKAN-Ivashkov-etal, QKAN-variation, QKAN-enh-var}.
However, there is a gap between achieving quantum theoretical results and describing quantum analogues of the classical Kolmogorov--Arnold representation theorem, which is best seen in Fig.~\ref{fig:qka-gap}.
Some attempts to fill this gap were made, for example, in \cite{Dzhenzher26-alg-QKA, Dzhenzher26-unit-QKA}.
In this work, we continue these efforts, this time in the setting of unitary evolutions over coherent states.

So, in this paper, we consider a fixed-input-state model: the dependence on the argument \(x\) is contained entirely in the supplied copies of the coordinate qubits, whereas the unitary operator depends on the target function and the number of copies, but not on \(x\). Our additional requirement is a Kolmogorov--Arnold structure with universal inner observables and a single outer function, common to all approximation levels. Applying a Sprecher-type representation to the rotation angle \(\arcsin\sqrt f\), we obtain a factorisation into \(2d+1\) commuting unitary factors. The resulting output probability equals the Bernstein polynomial \(B_nf\) exactly, while, for continuous \(f\), the joint output converges uniformly to the product of the original input and a qubit encoding \(f(x)\). The result is a representation and approximation theorem; no bound on the elementary-gate complexity is asserted.

The structure of the article is as follows.
Section~\ref{s:main} is devoted to the theoretical background and the formulation of the main result (Theorem~\ref{t:qbka}).
Section~\ref{s:proof} provides the proof of the main result.
Section~\ref{s:discuss} gives some comments on the proof of the result and its optimisation.
Finally, Section~\ref{s:conclusion} concludes the paper with a discussion of further research.

\section{Background and the main result}\label{s:main}

The Kolmogorov--Arnold representation theorem has been optimised in many ways \cite{Hedberg-appx, Lorentz-Golitschek-Makovoz, Sprecher1965}.
After that, the constructive proof was provided \cite{Sprecher96, Sprecher97, Braun-Griebel-2009}.
We will use the following generalisation from \cite{Ismailova-Ismailov}.

\begin{theorem}[Ismailova, Ismailov; 2024]\label{t:ismailovs}
    Let $d>1$ be an integer.
    There exist a universal increasing H\"older function \(\varphi\colon\R\to\R\) and constants \(\alpha,\lambda_1,\ldots,\lambda_d\in\R\), such that \\
    for any ``target'' $d$-variate function \(f\colon[\,0,1\,]^d\to\R\), \\
    there exists an ``outer'' univariate function \(g\colon\R\to\R\), such that
    \[
        f(x_1,\ldots,x_d) = \sum_{q=0}^{2d}g\left(\sum_{p=1}^d \lambda_p\varphi(x_p+\alpha q) + (2d+1)q\right).
    \]
\end{theorem}

Note that Sprecher \cite{Sprecher1965, Sprecher96, Sprecher97} showed this representation for continuous functions.
Here, the function $f$ can be made discontinuous. So, if $f$ is continuous, discontinuous, or unbounded, then so is $g$.
Also, all the constants \(\alpha,\lambda_p\) can be easily computed, and the function $\varphi$ is defined constructively.
Its H\"older constant is also controllable.

Recall that \(\ket{0}\) and \(\ket{1}\) are basis vectors of a $1$-qubit state.
For $t\in[\,0,1\,]$, let us consider the coherent states
\[
    \ket{t} := \sqrt{1-t}\ket{0} + \sqrt{t}\ket{1}.
\]
It may seem strange that \(\sqrt{t}\) does not stand near \(\ket{0}\), but it is just to avoid contradictions:
\[
    \ket{0} = \sqrt{1-0}\ket{0} + \sqrt{0}\ket{1}.
\]

We presume that we can encode our input in the following states: for any integer \(n\geq 1\) and a point \(x = (x_1,\ldots, x_d)\in[\,0,1\,]^d\), define
\[
    \ket{\Psi_n(x)} := \bigotimes_{p=1}^d \ket{1-x_p}^{\otimes n}.
\]
We presume that these input states are given to us.
It is clear that if we know the precise $x_p$, we can prepare such states.

It is well known \cite{NacuPeres-fast-bern} that the Bernstein polynomials
\[
    B_nf(x_1,\ldots,x_d) := 
    \sum_{k_1,\ldots,k_d=0}^n f\left(\frac{k_1}{n},\ldots,\frac{k_d}{n}\right)\prod_{p=1}^d\binom{n}{k_p}x_p^{k_p}(1-x_p)^{n-k_p}
\]
uniformly approximate a continuous $d$-variate function $f$.
Our idea is to use the Kolmogorov--Arnold representation not for $f$ but for $B_nf$, thus approximating $f$ as well.
Strictly speaking, we will obtain $B_nf$ as the probability of measuring unity in the last qubit.

Recall that
\[
    \Ybf = \begin{pmatrix}
        0 & -i \\ i & 0
    \end{pmatrix}
\]
is the Pauli matrix on a one qubit, and \(\Ibf\) is the identity operator (not necessary on a one qubit).

Our main result is as follows.

\begin{theorem}\label{t:qbka}
    \begin{enumerate}[label=(\roman*), ref=\roman*]
    \item\label{e:bern-poly}
    Let $d>1$ be an integer.
    There exist universal Hermitian operators
    \[
        \Hbf_q^{(n)},
        \qquad q=0,\ldots,2d, \quad n\geq 1,
    \]
    such that for any ``target'' $d$-variate function \(f\colon[\,0,1\,]^d\to[\,0,1\,]\), \\
    there exists an ``outer'' univariate function \(g\colon\R\to\R\) (independent of $n$) \\ as well as ``outer'' unitary operators
    \[
        \Ubf_n := \prod_{q=0}^{2d} \exp\left(-i g(\Hbf_q^{(n)})\otimes \Ybf\right),
    \]
    such that the probability \(p_n(x)\) of obtaining the unit in the ancilla qubit equals $B_nf(x)$:
    \[
        p_n(x) := \abs{(\Ibf \otimes \bra{1})\Ubf_n(\ket{\Psi_n(x)}\otimes\ket{0})}^2 = B_nf(x).
    \]
    Moreover, if $f$ is continuous, then so is $g$.
    
    \item\label{e:coh} Additionally,
    \begin{multline*}
        d_n(x) := \norm{\Ubf_n \ket{\Psi_n(x)}\otimes\ket{0} - \ket{\Psi_n(x)}\otimes\ket{f(x)}}^2 = \\
        = 2 - 2 \left(\sqrt{f(x)}B_n(\sqrt{f})(x) + \sqrt{1-f(x)}B_n\sqrt{1-f}(x)\right).
    \end{multline*}
    Moreover, if $f$ is continuous, then \(d_n \xrightarrow[n\to\infty]{}0\) uniformly on $x$.
    \end{enumerate}
\end{theorem}

Note that the theorem holds even for discontinuous functions $f$.
It does not provide the approximation in this case, but it may yield additional estimates based on the properties of Bernstein polynomials.
For example, if $f$ has all $2^d$ left, right, upper, lower, etc. limits for some $x$, then \(B_nf(x)\) converges to their average, and, by the theorem, so $p_n$ does.

\section{Proof of Theorem~\ref{t:qbka}}\label{s:proof}

% \begin{proof}[Proof of Theorem~\ref{t:qbka}]
    Let \(\Nbf_p\) be a zero number operator for $n$ qubits in the $p$-th set.
    That is,
    \[
        \Nbf_p = \Nbf_p^{(n)} := \sum_{k=1}^n \Ibf^{\otimes (p-1)n+k-1} \otimes \ket{0}\bra{0} \otimes \Ibf^{\otimes (d-p+1)n-k}.
    \]
    Let
    \[
        \Xbf_p = \Xbf_p^{(n)} := \frac{\Nbf_p^{(n)}}{n}
    \]
    be a zero fraction operator.

    Take the universal function \(\varphi\colon\R\to\R\) and constants \(\alpha,\lambda_p\in\R\) from Theorem~\ref{t:ismailovs}.
    Put
    \[
        \Hbf_q = \Hbf_q^{(n)} := \sum_{p=1}^d \lambda_p\varphi(\Xbf_p+\alpha q) + (2d+1)q.
    \]
    Applying Theorem~\ref{t:ismailovs} for the target function
    \[
        \theta(x) := \arcsin\sqrt{f(x)},
    \]
    obtain the outer function $g$.
    So,
    \[
        \theta(\Xbf_1,\ldots,\Xbf_d) = \sum_{q=0}^{2d}g\left(\Hbf_q\right).
    \]
    Since all \(\Hbf_0,\ldots,\Hbf_{2d}\) commute, we have
    \[
        \Ubf_n = \exp\left(-i\theta(\Xbf_1,\ldots,\Xbf_d) \otimes \Ybf\right).
    \]
    On the eigenspace corresponding to \(k_1,\ldots,k_d\) zeros in the $p$-th $n$-qubit sets, \(\theta(\Xbf_1,\ldots,\Xbf_d)\) acts like multiplication by \(\theta(\frac{k_1}{n},\ldots,\frac{k_d}{n})\).
    Thus, on these eigenspaces, \(\Ubf_n\) acts on the last qubit like
    \[
        \exp\left(-i\theta\left(\frac{k_1}{n},\ldots,\frac{k_d}{n}\right) \Ybf\right) =
        \cos \theta - i \Ybf \sin \theta,
    \]
    which, with \(\theta(x) = \arcsin \sqrt{f(x)}\), gives
    \[
        \exp(-i\theta \Ybf)\ket{0} = \cos \theta \ket{0} + \sin \theta \ket{1} = \Ket{f\left(\frac{k_1}{n},\ldots,\frac{k_d}{n}\right)}.
    \]

    Now, for \(b \in \{0,1\}^{dn}\), let \(k_p = k_p(b)\) be the number of zeros in the $p$-th $n$-group.
    Denote
    \[
        c_b(x) := \prod_{p=1}^d x_p^{\frac{k_p(b)}{2}}(1-x_p)^{\frac{n-k_p(b)}{2}}.
    \]
    Then,
    \[
        \ket{\Psi_n(x)} = \sum_{b \in \{0,1\}^{nd}} c_b(x)\ket{b}
    \]
    (here we identified \(\ket{b}\) with \(\bigotimes_{s=1}^{nd} \ket{b_s}\).)
    Since
    \[
        \Xbf_p \ket{b} = \frac{k_p(b)}{n}\ket{b},
    \]
    we obtain
    \[
        \Ubf_n \ket{\Psi_n(x)}\otimes\ket{0} = \sum_{b\in\{0,1\}^{nd}} c_b(x)\ket{b}\otimes \Ket{f\left(\frac{k_1(b)}{n},\ldots,\frac{k_d(b)}{n}\right)}.
    \]
    Hence
    \[
        (\Ibf \otimes \bra{1})\Ubf_n(\ket{\Psi_n(x)}\otimes\ket{0}) = \sum_{b\in\{0,1\}^{nd}} c_b(x)\sqrt{f\left(\frac{k_1(b)}{n},\ldots,\frac{k_d(b)}{n}\right)}\ket{b},
    \]
    and
    \[
        p_n(x) = \sum_{b\in\{0,1\}^{nd}} \abs{c_b(x)}^2f\left(\frac{k_1(b)}{n},\ldots,\frac{k_d(b)}{n}\right) = B_nf(x).
    \]
    Thus, \eqref{e:bern-poly} is proved.
    
    In order to prove~\eqref{e:coh}, for fixed \(x=(x_1,\ldots,x_d)\in[\,0,1\,]^d\),
    let \(\xi_1,\ldots,\xi_d\) be independent binomial random variables with distributions \(\xi_p\sim\Bin(n,x_p)\).
    Since in the decomposition of \(\Ubf_n \ket{\Psi_n(x)}\otimes\ket{0}\), the coefficients \(c_b(x)\) represent the probabilities of exactly the same independent distributions, and all the states \(\ket{b}\) are pairwise orthogonal, we obtain that
    Then
    \[
        d_n(x) = \Expect \norm{\Ket{f\left(\frac{\xi_1}{n},\ldots,\frac{\xi_d}{n}\right)} - \ket{f(x)}}^2.
    \]
    For any \(s,t\in[\,0,1\,]\),
    \[
        \norm{\ket{s} - \ket{t}}^2 = \left(\sqrt{s}-\sqrt{t}\right)^2 + \left(\sqrt{1-s} - \sqrt{1-t}\right)^2 =
        2 - 2\left(\sqrt{st} + \sqrt{(1-s)(1-t)}\right).
    \]
    Hence, substituting in the formula above \(s= f(x)\) and \(t = f\left(\frac{\xi_1}{n},\ldots,\frac{\xi_d}{n}\right)\), we obtain
    \[
        d_n(x) = 2 - 2 \sqrt{f(x)} \Expect \sqrt{f\left(\frac{\xi_1}{n},\ldots,\frac{\xi_d}{n}\right)} - 2 \sqrt{1-f(x)} \Expect \sqrt{1-f\left(\frac{\xi_1}{n},\ldots,\frac{\xi_d}{n}\right)}.
    \]
    It remains to be noted that these expected values are precisely \(B_n\sqrt{f}(x)\) and \(B_n\sqrt{1-f}(x)\), respectively.
% \end{proof}

\section{Discussion of the estimates}\label{s:discuss}

In this paragraph, we show that in Theorem~\ref{t:qbka}, we need at least $n$ copies of each state \(\sqrt{x_p}\ket{0}+\sqrt{1-x_p}\ket{1}\).
Indeed, let
\[
    f(x) := e^{-d}e^{\sum_{p=1}^d x_p}
\]
(here $e^{-d}$ is just the norming coefficient).
Then
\[
    B_nf(x) =
    e^{-d}\sum_{k_1,\ldots,k_d=0}^n e^{\frac{k_1}{n}+\ldots+\frac{k_d}{n}}\prod_{p=1}^n\binom{n}{k_p}x_p^{k_p}(1-x_p)^{n-k_p} =
    e^{-d}\prod_{p=1}^d \left(x_pe^{1/n} + (1-x_p)\right)^n,
\]
which shows that \(x_p\) stands in the power of $n$ in \(B_nf(x)\).
So, in order to obtain it in \(p_n\), we need at least $n$ copies of the states \(\sqrt{x_p}\ket{0}+\sqrt{1-x_p}\ket{1}\).
See some analogous estimates for a quantum polynomial approximation in, for example, \cite{Beals-quant-poly}.

However, there is another way to optimise the result.
As we can see from the proof, in the decompositions
\[
    \ket{\Psi_n(x)} = \sum_{b \in \{0,1\}^{nd}} c_b(x)\ket{b}
\]
and
\[
    \Ubf_n \ket{\Psi_n(x)\otimes\ket{0}} = \sum_{b\in\{0,1\}^{nd}} c_b(x)\ket{b}\otimes \Ket{f\left(\frac{k_1(b)}{n},\ldots,\frac{k_d(b)}{n}\right)},
\]
we really need to identify the states for which \(k_i(b)=const_i\).
So, considering the Dicke states
\[
    \ket{D_{n,k}} := \frac{1}{\sqrt{\binom{n}{k}}} \sum_{\substack{w\in\{0,1\}^n \\ \sum w_i=n-k}} \ket{w},
\]
we can rewrite
\[
    \ket{\Psi_n(x)} = \bigotimes_{p=1}^d \sum_{k=0}^n\sqrt{\binom{n}{k}x_p^k(1-x_p)^{n-k}}\ket{D_{n,k}}.
\]
Thus, instead of preparing \(dn\) qubit states, we can prepare only $d$ states in the space of Dicke states.
So, we are able to lower the dimension of the input state from \(dn\) to \(d \lceil\log_2(n+1)\rceil\).
Note that this compression of the input space does not prevent us from obtaining the probability \(p_n\), but it nevertheless leads to some information loss.
See some analogues of such compression in \cite{Hoch-quant-bern, Plesch-Buzek-compress}.

Finally, it is worth mentioning that Theorem~\ref{t:qbka} is equipped with all the delightful properties of the classical Bernstein approximation.
For example, if $f$ is a Lipschitz function, then both in the convergence of \(p_n\) to $f$ and of \(d_n\) to $0$, we not only obtain uniform convergence, but we can also estimate it using the Lipschitz constant.

\section{Conclusions and further research}\label{s:conclusion}

We established a quantum Bernstein--Kolmogorov--Arnold approximation theorem for functions on the unit cube. The construction transfers a classical superposition formula to the joint functional calculus of universal commuting observables and yields a product of $2d+1$ unitary factors. The dependence on the target function is contained in a single outer function, independent of the number of input copies and continuous whenever the target function is continuous. The output probability reproduces the corresponding Bernstein polynomial exactly. For continuous targets, the full output state converges uniformly to the tensor product of the original input state and a qubit encoding the target value. Thus, the result connects classical superposition theory with both probabilistic and coherent quantum approximation.

A natural direction for further research is to develop Quantum Kolmogorov--Arnold representation and approximation theorems for more general families of quantum states and operator-valued maps. In particular, one may ask which fixed families of inner observables admit universal representations of a prescribed class of targets. In the classical setting, analogous questions are closely related to basic embeddings and the geometric obstructions studied by Sternfeld~\cite{Sternfeld85,Sternfeld89}. Those results showed that the number $2d+1$ in the Kolmogorov--Arnold representation theorem is optimal; see an overview on it, for example, in \cite{Dzhenzher24}.
It would be interesting to formulate quantum counterparts of basic embeddings and to investigate whether analogues of Sternfeld arrays can provide criteria for universality or obstructions to it~\cite{MKT-2003}. Such a theory would complement constructive Quantum Kolmogorov--Arnold results with statements describing when representations of a given form are impossible. See, for example, \cite{Freedman25}, where some connection of Sternfeld array and quantum entourage is provided.

\printbibliography
% % \printbibitembibliography

\end{document}